# Weyl semimetals as catalysts

Catherine R. Rajamathi [1], Uttam Gupta [2], Nitesh Kumar [1], Hao Yang [4], Yan Sun [1], Vicky Süβ [1], Chandra Shekhar [1], Marcus Schmidt [1], Binghai Yan [1, 3], Stuart Parkin [4], Claudia Felser [1, *] and C.N.R. Rao [2, *]

[1] *Max Planck Institute for Chemical Physics of Solids, 01187 Dresden, Germany.*

[2] *Chemistry and Physics Materials Unit, New Chemistry Unit and International Centre for Materials Science, Sheik Saqr Laboratory, Jawaharlal Nehru Centre for Advanced Scientific Research, Jakkur P. O., Bangalore 560064, India*

[3] *Max Planck Institute for the Physics of Complex Systems, 01187 Dresden, Germany.*

[4] *Max Planck Institute for Microstructure Physics, 06120 Halle (Saale), Germany.*

*e-mail: felser@cpfs.mpg.de and cnrrao@jncasr.ac.in

**The search for highly efficient and low-cost catalysts is one of the main driving forces in catalytic chemistry. Current strategies for the catalyst design focus on increasing the number and activity of local catalytic sites, such as the edge-sites of molybdenum disulfides in the hydrogen evolution reaction (HER). Here, we propose and demonstrate a different principle that goes beyond local site optimization by utilizing topological electronic states to spur catalytic activity. For HER, we have found excellent catalysts among the transition-metal monopnictides - NbP, TaP, NbAs, and TaAs - which were recently discovered to be topological Weyl semimetals. In addition to considerations of free energy, we explore the role of metallicity, carrier mobility and topological electronic states in the remarkable HER performance of these materials. Here we show that the combination of robust topological surface states and large room temperature carrier mobility, both of which originate from bulk Dirac bands of the Weyl semimetal, are a recipe for high activity HER catalysts. Thus, our work provides a guiding**

**principle for the discovery of novel catalysts from the emerging field of topological materials.**

Heterogeneous catalysis plays a major role in organic and inorganic chemistry. The high complexity of the catalytic process is a major challenge.[1-4] One important catalytic process, of increasing importance, is the harvesting of solar energy to produce hydrogen ($H_2$) from water: one of the major challenges in the supply of "green energy".[5,6] The educts, products and the surface of the catalyst have to be taken into account in heterogeneous catalysis. Many conventional metals and semiconductors are under investigation as potential catalytic materials.[7-9] Here we introduce a new concept based on the use of unconventional topological materials as catalysts. A stable supply of itinerant electrons at the surface is advantageous for catalytic processes involving redox reactions. Materials with a high mobility of electrons and holes reduces the probability of recombination of the electron-hole pairs that are created in the redox process. High carrier mobilities that are realized from the linear bands of a Dirac cone are a fundamental property of Weyl- and Dirac-semimetals.[10,11] Whilst these unusual semimetals have drawn significant attention from the condensed matter community, their unique properties have not yet been considered by chemists. The problem of surface contamination, which is a bottleneck in the field of catalysis, can be diminished to a great extent in topological materials because of their robust topologically-protected surface states.[12,13]

The precondition for Weyl semimetals, which are topological insulators (TI), is a band inversion. Many compounds containing heavy metals show such a band inversion, more commonly referred to, within the chemistry community, as an inert pair effect.[14] In a relativistic band structure, band crossing is forbidden such that, in topological insulators, a new band-gap opens, and a surface-state with a Dirac cone electronic structure appears. At the borderline between trivial and topological insulators Dirac and Weyl semimetals appear.

In a Weyl semimetal, pairs of Dirac cones are formed in the bulk of the material where the number of pairs depends on the detailed symmetry of the particular metal. Weyl semimetals also exhibit unusual surface states with open Fermi arcs. To verify our hypothesis that the new topological materials with superior electronic properties can also be highly efficient in catalysis we have tested them for dye-sensitized hydrogen evolution reaction (HER). In the HER, solar light is absorbed by photon capture systems, such as the dye Eosin Y (EY). The resultant excited electrons can be transferred to the catalyst, on which $H^+$ in the water is reduced to form $H_2$. A schematic representation of the HER process is shown in Fig. 1.

Among photocatalysts for hydrogen production, $MoS_2$ nanoparticles were reported to demonstrate high efficiency to catalyze both photochemical as well as electrochemical HER.[15-17] The metallic form of $MoS_2$ either in the form of metallic edges of 2H-$MoS_2$ [16] or the metastable 1T polytype [16] is found to be much more active than its semiconducting counterpart. Recent theoretical studies also predicts metallic transition metal dichalcogenides (TMDs) to be excellent candidates for HER.[18] However, our proposal goes beyond these proposed advantages of metallic dichalcogenides since the 1T polytypes of the $MoX_2$ family are also topologically non-trivial. We are convinced that the electronic structure of Weyl semimetals favor catalysis. To test whether the metallicity alone is sufficient or rather it is the topology that is at work, we have compared the catalytic performance of several different TMDs for HER: topologically trivial semiconducting 2H-$MoTe_2$; topologically trivial metallic 1T-$TaS_2$; and topologically nontrivial 1T'-$MoTe_2$. Note that Ta has a $d^1$ configuration (octahedral coordination) in 1T-$TaS_2$ and Mo has a $d^2$ configuration (trigonal prismatic coordination) in 2H-$MoS_2$, which makes the first system metallic and the latter semiconducting [11]. We have chosen 1T-$TaS_2$ for comparison since it was predicted to be an active and stable metallic transition metal catalyst.[18]

Calculated band structures of 2H-MoTe$_2$, 1T'-MoTe$_2$ and 1T-TaS$_2$ are compared in Figs 2a-c. We find that 2H-MoTe$_2$ is a large band gap material whereas MoTe$_2$ in the 1T' phase, which is a distorted 1T structure, behaves as a semimetal, in agreement with transport measurements.[19] To understand the role of inherent metallicity and nontrivial electronic structure, we compare the HER activity of 1T'-MoTe$_2$, 2H-MoTe$_2$ and 1T-TaS$_2$ in Fig. 2d. 1T'-MoTe$_2$ exhibits a huge H$_2$ evolution reaching a value of 1294 μmoles g$^{-1}$ in 6 hours. However, semiconducting 2H-MoTe$_2$ evolves only 28.14 μmoles g$^{-1}$ in the same period. Surprisingly, metallic 1T-TaS$_2$ is completely inactive. In Fig. 2e, we show a histogram of the rate of H$_2$ evolution of the three compounds where we can clearly see that 1T'-MoTe$_2$ is the most active catalyst. The Gibb's free energy ($\Delta G_{H*}$) of adsorption of hydrogen at the catalyst surface has been recognized as the best thermodynamic parameter to define the activity of an HER catalyst. The closer this value is to zero the better is the performance. The $\Delta G_{H*}$ (on abscissa) and the activity (on ordinate) hence make a so-called volcano diagram (Fig. 3b). In addition to the fact that the metallic nature of 1T'-MoTe$_2$ facilitates the HER process compared to the semiconducting 2H-MoTe$_2$, a relatively larger $\Delta G_{H*}$ in 2H-MoTe$_2$ also explains its smaller activity.

Notwithstanding that 1T-TaS$_2$ and 1T'-MoTe$_2$ are metallic with comparable $\Delta G_{H*}$ values, 1T'-MoTe$_2$ exhibits topological features in its band structure, encouraging us to consider topological effects. The conduction and valence bands are inverted for the single layer of 1T'-MoTe$_2$, causing a TI[20] where metallic edge states exist at the boundary.[12,13] Such an inverted band structure further leads to a three-dimensional (3D) topological Weyl semimetal (TWS) in Td-MoTe$_2$,[21] which originates from the slightly distorted 1T'-MoTe$_2$ phase by small strain or temperature reduction.[19]

In a Weyl semimetal, the conduction and valence bands cross each other linearly through nodes, called Weyl points, near the Fermi energy. As a 3D analogue of graphene, TWSs are expected to exhibit very high mobility in their charge transport.[11] Similar to TIs, TWSs also present robust metallic surface states[22] that are stable against defects, impurities, and other surface modifications. Analogous to the role of graphene, in the $MoS_2$ catalyzed HER, we believe that the highly mobile TWS bulk states help electrons diffuse freely and quickly. Further, the topological surface states may cause the surface to act as stable active planes for catalysis. The first family of TWSs was experimentally discovered in transition metal monopnictides: NbP, TaP, NbAs, and TaAs, by revealing the topological surface states.[23-27] These materials are semimetals wherein Weyl points are located near the Fermi level with a total of 12 pairs of Weyl nodes in the first Brillouin zone (Figs 3a). Therefore, we investigated the HER activity in these TWS compounds.

The HER activities of NbP, TaP, NbAs, and TaAs were studied over a period of 6 hours. Our studies show that all four TWSs are highly active (Fig. 3c) and NbP being the lightest among all, performs the best as an HER catalyst with the highest value of $H_2$ evolved per gram of the catalyst (3520 $\mu$moles $g^{-1}$). The compounds can undergo many cycles of HER without activity fading as can be seen in Fig. 3b where we show three cycles of HER in NbP with comparable performance each time. As discussed in the Supplementary Information no change in composition was found after several HER cycles. We show the activity and turnover frequency (TOF: the number of moles of $H_2$ evolved per mole of catalyst used) as histograms of all four compounds in Fig. 3d. In general, phosphides are better HER catalysts than arsenides. This trend is well captured by the relative $\Delta G_{H*}$ values of these compounds. NbP has the lowest $\Delta G_{H*}$ among all followed by TaP, TaAs and NbAs and interestingly TOF also follows the same trend.

Having investigated the thermodynamical aspects of the catalysts we now focus on the role of kinetics. As we know that the reduction of water occurs at the surface of the catalyst, increasing the surface area of the catalyst should result in increased activity of the catalyst. For this we have selected NbP as an example and compared the activity in single crystals crushed into powder and polycrystalline material obtained by solid state reaction. We encounter a two fold increase in the activity of polycrystals with larger surface area compared to the single crystals. The activity in terms of per gram of the catalyst and TOF is comparable to catalysts like Ni metal nanoparticles[28,29] and highly active platinum decorated $TiO_2$ nanoparticles, under similar experimental conditions.

In order to draw any conclusive effect of the kinetics we must scale the activity per surface area of the catalyst (Fig. 4d). Interestingly, here NbP performs much better than the Ni nanoparticles with an activity that is one order of magnitude higher despite the fact that the latter has a $\Delta G_H$ value closer to zero. Moreover, the HER activity of NbP is also higher compared to Pt-$TiO_2$ (Pt-P25) where the catalytic sites mostly reside at the metallic surface of Pt. The titania nanoparticles used were a mixture of anatase and rutile and the interfaces of these two polymorphs have been identified as an excellent medium for electron and hole separation. For $SrNbO_3$, a well-known visible light absorbing metallic oxide, the activity per unit surface area is two orders of magnitude smaller than for NbP.[30]

Having considered the fact that the value of $\Delta G_{H*}$ is more favorable for Ni than NbP, we now compare their electronic properties in order to gain insights into the higher HER activity of NbP. We note that Ni is highly metallic with a room temperature conductivity of ~ $10^7$ S/m; on the other hand NbP is semimetallic (~ $10^6$ S/m at room temperature). As the conductivity σ is related to mobility through the density of charge carriers (σ = μne, where μ is mobility, n is the carrier density, e is the electronic charge), the carriers are much more mobile in NbP as compared to Ni, because of the much smaller carrier density in NbP. The average mobility of NbP is of the order of $10^3$ cm$^2$/Vs as compared to ~10 cm$^2$/Vs in Ni at room temperature.[11] The effect of mobility on the hydrogen evolution reaction has been discussed in the literature but, however, mostly focused on composite catalysts where graphene is used as a medium to provide a high mobility of the carriers.[31] The large mobilities facilitates the rapid transfer of carriers for the catalytic reactions, thereby enhancing the kinetics. It also helps for the rapid separation of electrons and holes. Further research is needed to accurately identify the active sites involved in the HER for the NbP family of compounds. However, recent STM studies on TaAs shows that the bulk Weyl nodes and, therefore, the states close to the projected Fermi arcs on the surfaces predominantly carry Ta-orbital character.[32] This implies that the As-states as well as the surface states from the trivial bands are prone to delocalize into the bulk whereas the Ta states are comparatively more robust on the surface. We therefore speculate an important role for transition metal states on the hydrogen evolution activity in our catalysts.

TWSs serve as excellent candidates for catalysis. We have shown that inherent metallicity alone in a material does not improve catalytic activity and that the most important factors that one must consider are electronic properties as well as the mobility of carriers. Among the transition metal monopnictides, the phosphides are better for HER compared to the arsenides. These catalysts are also robust and therefore can be used long-term. They are also durable and have no significant changes in chemical composition after the catalytic procedure. The activities of TWSs should be improved significantly by reducing the particle size.

To conclude, we have shown for two distinct families of Weyl semimetal that their topological electronic structure strongly influences the catalysis of the redox reaction with regard to H$_2$ evolution from water. We have identified 1T'-MoTe$_2$ and NbP and related monopnictides as excellent catalysts for the H$_2$ evolution reaction. Our findings suggest that topological metals or semimetals with high mobilities, robust surface states and a stable supply of itinerant electrons are promising candidates for catalysis, and thus provides a new route to the discovery of efficient catalysts. Moreover, we propose that the chiral surface state of topological materials may even pave the path for asymmetric catalysis.

**Methods**

1T′-MoTe$_2$ and 2H-MoTe$_2$ crystals were grown via chemical vapour transport using polycrystalline MoTe$_2$ powder and TeCl$_4$ as a transport additive [22]. 1T-TaS$_2$ was grown from its elements via solid state route. Stoichiometric quantities of Ta (Alfa Aesaer 99.999%) and S (Alfa Aesar 99.99%) were sealed in a quartz ampoule and heat treated at 900°C for one week and quenched. Single crystals of NbP, TaP, NbAs and TaAs were grown via chemical vapour transport. Stoichiometric quantities of Nb (Alfa Aesar, 99.99%) and P (Alfa Aesar, 99.999%) were weighed in accurately in a quartz ampoule, flushed with Ar, sealed under vacuum and tempered in two consecutive steps of 600 °C and 800 °C for 24 h each prior to crystal growth. Crystal growth was carried out in a two-zone furnace between 850 – 950 °C for 2 weeks using I$_2$ (8 mol/ml) as the transport agent. Stoichiometric quantities of Nb (Alfa Aesar, 99.99%) and As (Chempur, 99.9999%) were tempered in a similar manner prior to crystal growth. Crystal growth was carried out in a two-zone furnace between 900 – 1000 °C for 4 weeks using I$_2$ (8 mol/ml) as the transport agent. Stoichiometric quantities of Ta (Alfa Aesar, 99.97%) and P (Alfa Aesar, 99.999%) were tempered in a similar manner prior to crystal growth. Crystal growth was carried out in a two-zone furnace between 900 – 1000 °C for 2 weeks using I$_2$ (12 mol/ml) as the transport agent. Stoichiometric quantities of Ta (Alfa Aesar, 99.97%) and As (Chempur, 99.9999%) were also tempered in a similar manner prior to crystal growth. Crystal growth was carried out in a two-zone furnace between 900 – 1000 °C for 4 weeks using I$_2$ (12 mol/ml) as the transport agent. The crystals obtained have been characterized by powder x-ray diffraction using an image-plate Huber G670 Guinier camera with a Ge (111) monochromator and CoKα radiation. Indexing was done with the program WINX POW19 and PowderCell.20. The diffraction pattern is in good agreement with the calculated pattern. The FESEM images were recorded using Nova Nano FESEM 600 FEI.

The single crystals of Weyl semimetals (NbP, NbAs, TaP, TaAs, MoTe$_2$) were dispersed in 48 mL of triethanolamine (15% v/v) aqueous solution and sonicated for 30 minutes to ensure particle dispersion. 0.014 mmole of dye was added to the flask. The vessel was placed on a magnetic stirrer and thoroughly purged with N$_2$ gas for 5 minutes to free the surface of the catalyst by previously adsorbed gas and also to remove dissolved $O_2$ from the solution. The flask was then sealed with a rubber septum and irradiated with 100 W halogen lamp with constant stirring. The evolved gases were manually collected from the headspace of the vessel and analysed with a thermal conductivity detector TCD in PerkinElmer Clarus ARNEL 580GC gas chromatograph.

Electronic structures were calculated by the density functional theory (DFT) method as implemented in the Vienna *ab initio* Simulation Package (VASP). The exchange–correlation was considered in the revised Perdew-Burke-Ernzerhof (rPBE) parameterized generalized gradient approximation (GGA)[33] and spin–orbital coupling (SOC) was included. While calculating the adsorption energies, the van der Waals interaction using DFT-D2 method of Grimme[34] was included.

## Acknowledgments

This work was financially supported by ERC Advanced Grant No. (291472) 'Idea Heusler'.

## Author contributions

C.R.R., U.G., V.S., C.S., N.K., M.S., B.Y., S.S.P.P., C.F., and C.N.R.R. designed the experiments, carried out the experimental work and interpreted the experimental data. C.F and B.Y. proposed the concept of using TSMs for catalysis. B.Y. performed the theoretical calculations and analyzed the results. C.R.R., C.S., N.K., and S.S.P.P. coordinated the writing of the manuscript with input from all authors.

## Competing financial interests

The authors declare no competing financial interests.

**Figure captions**

**Figure 1 | Schematic of a topological Weyl semimetal (TWS) for catalyzing the dye-sensitized hydrogen evolution**. When light falls upon Eosin Y dye (EY) it is excited and in the presence of the sacrificial agent, triethanolamine (TEAO), the dye transfers an electron to the surface of the TWS, leading to charge separation, and, thereby, reducing water to hydrogen. In a TWS, the bulk bands are gapped by spin-orbit coupling (SOC) in 3D momentum space, except for some isolated linear- crossing points, namely Weyl points/ Dirac point.

**Figure 2 | Electronic band structure of several transition metal dichalcogenides (TMDCs) and rate of hydrogen evolution in HER.** Electronic band structure of **a,** semiconducting 2H-MoTe$_2$. **b,** semimetallic 1T'-MoTe$_2$ **c,** metallic 1T-TaS$_2$. **d,** Comparison of the rate of H$_2$ evolution using 1T'-MoTe$_2$, 2H-MoTe$_2$, and 1T-TaS$_2$. 1T'-MoTe$_2$ shows a higher activity compared to its 2H polymorph. (e) Histogram for the rate of H$_2$ evolution rate using 1T'-MoTe$_2$, 2H-MoTe$_2$, and 1T-TaS$_2$. The rate for polycrystalline 1T'-MoTe$_2$ is also included for comparison.

**Figure 3 | Electronic band structure of topological Weyl semimetals and its hydrogen evolution in catalysis. a,** Schematic band structure of transition metal monopnictides TWS family revealing semimetallic character. Weyl nodes of opposite chiralities are marked with blue and red dots  **b,** Predicted relative activities of various HER catalysts following the volcanic scheme as a function of calculated free energy of adsorption of hydrogen from the surface of the catalyst. **c,** Comparison of hydrogen evolution activity of different TWSs

(NbP, TaP, NbAs, and TaAs) powdered single crystals with an intermediate dye addition. **d,** Histogram of hydrogen evolution rate and TOF shown on left and right axes, respectively for all the four compounds.

**Figure 4 | $H_2$ evolution of NbP polycrystalline and single crystalline powders and their particle size before and after $H_2$ evolution. a,** Comparison of $H_2$ evolution of NbP in polycrystalline and single crystalline powder forms. Polycrystalline powders show higher catalytic activity compared to NbP single crystals. **b,** Cycling studies of polycrystalline NbP powder indicating the stability of $H_2$ evolution. **c**, Histograms for the rate of HER and TOF shown on the left and right axes, respectively for polycrystalline NbP, Ni metal nanoparticles, Pt-$TiO_2$ nanoparticles and $Sr_{0.9}NbO_3$. **d**, corresponding activity scaled to the surface area of the catalysts.

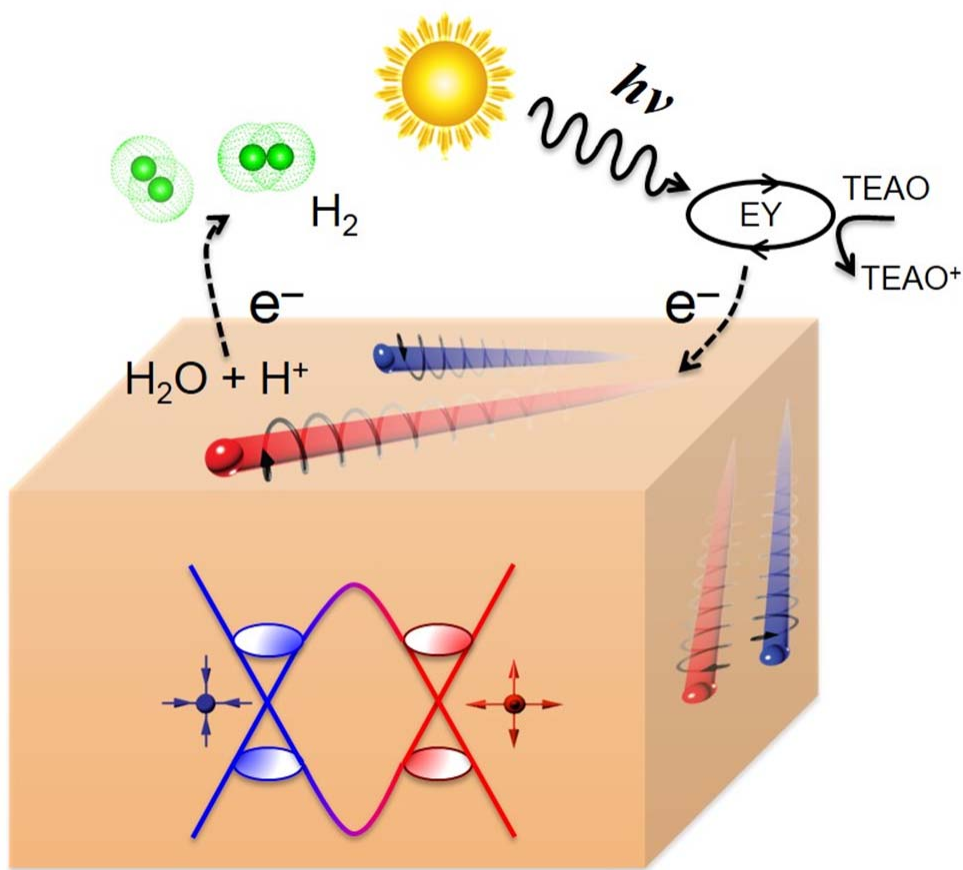

**Figure 1**

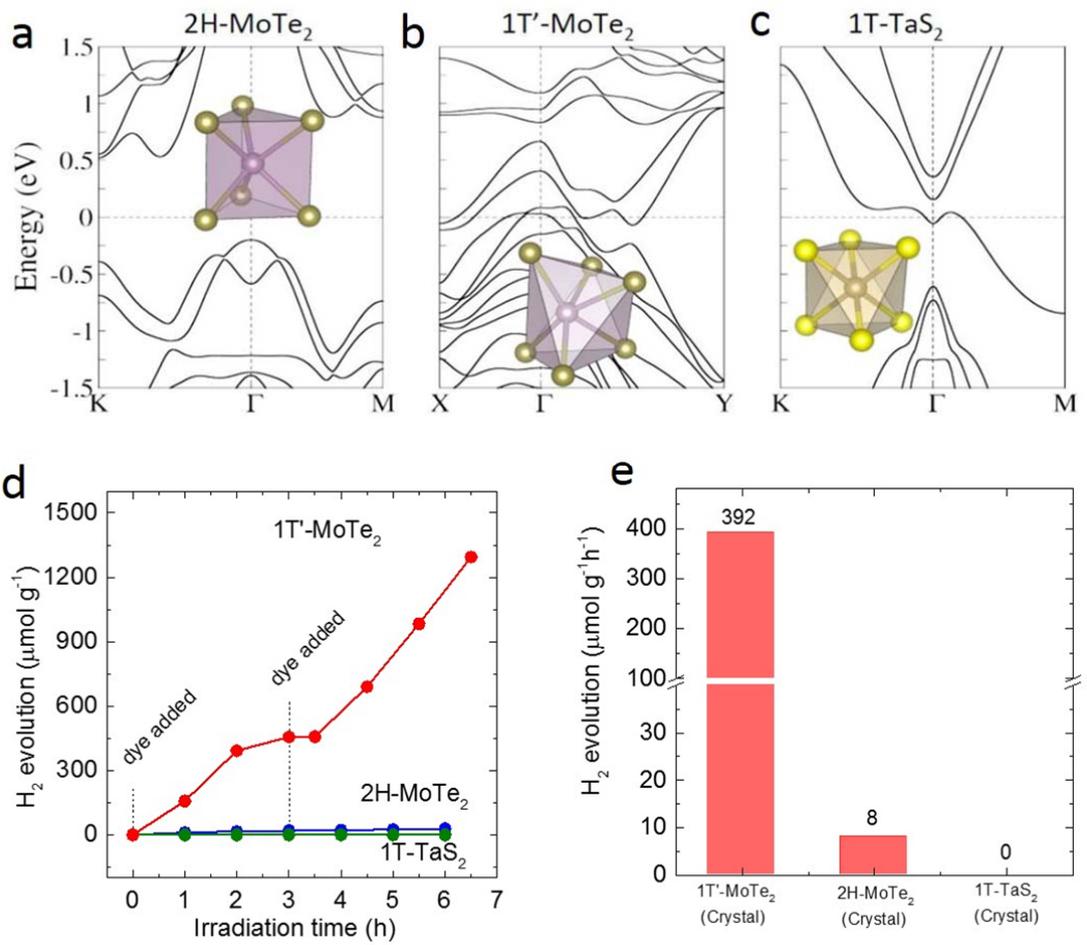

**Figure 2**

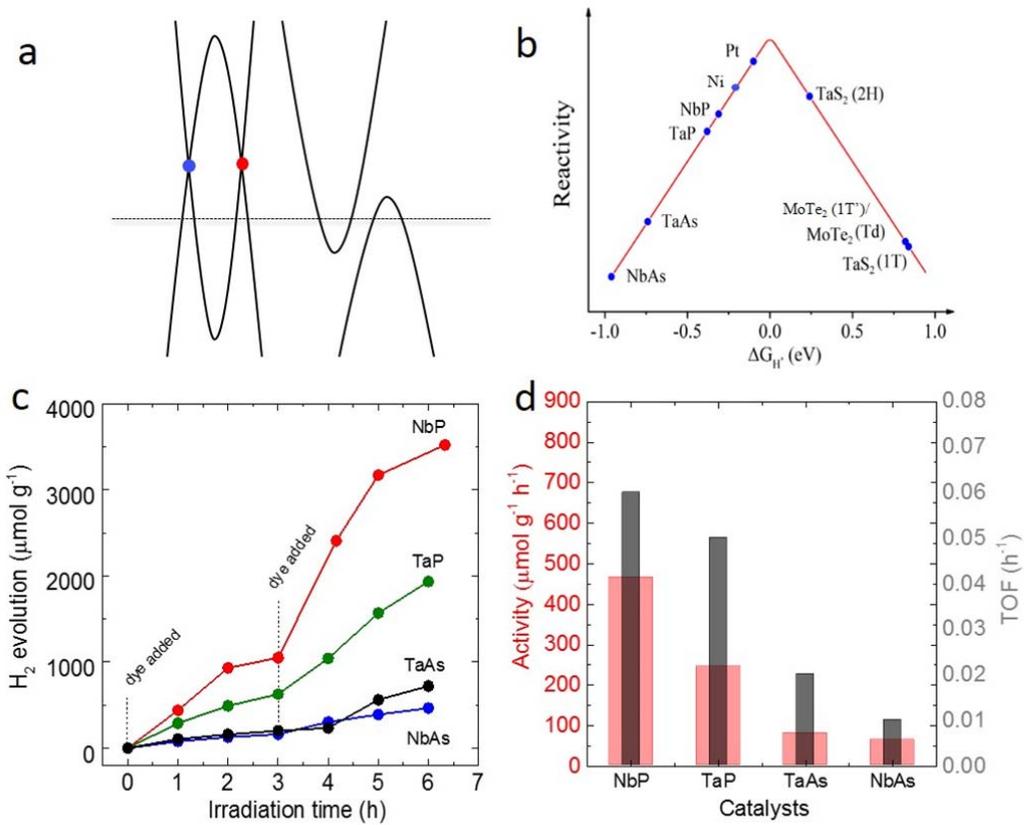

**Figure 3**

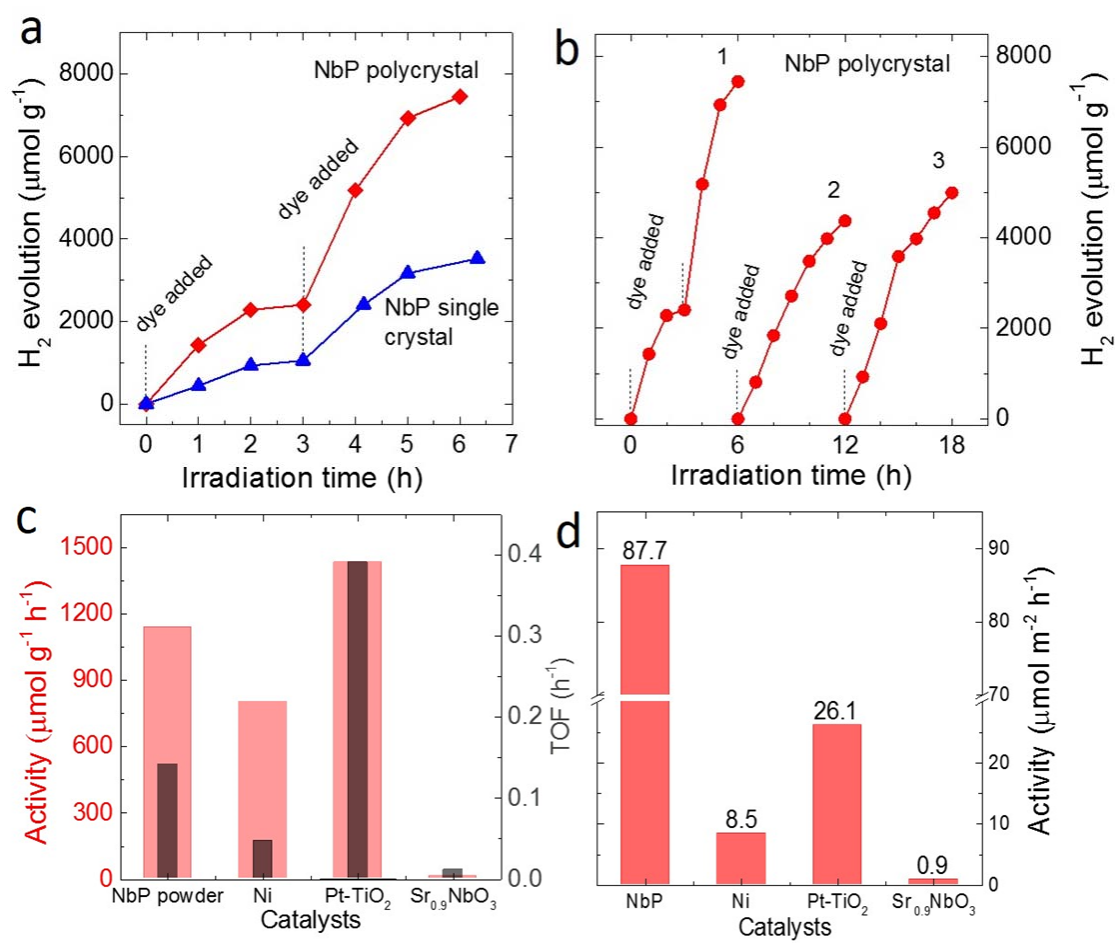

**Figure 4**

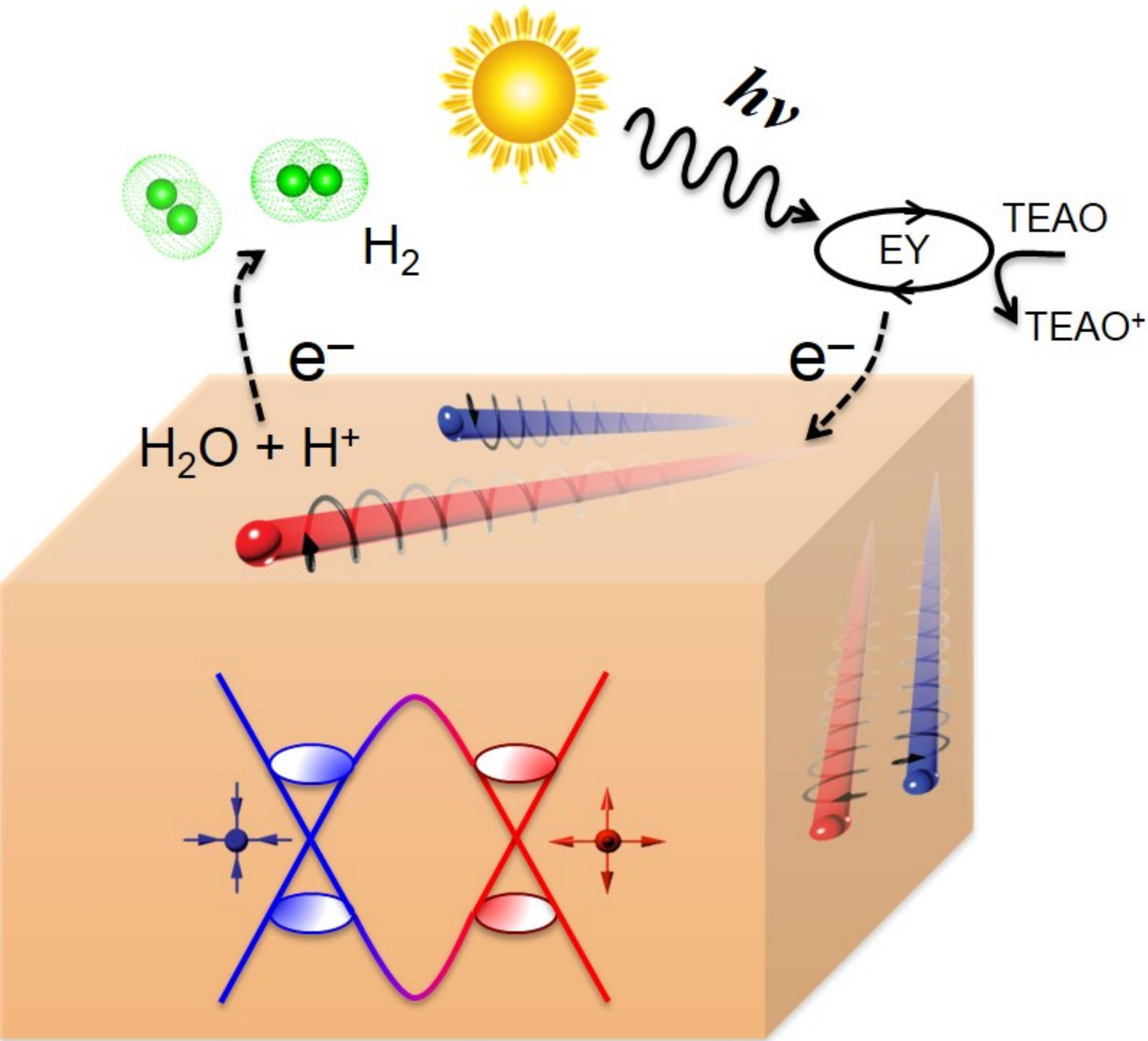

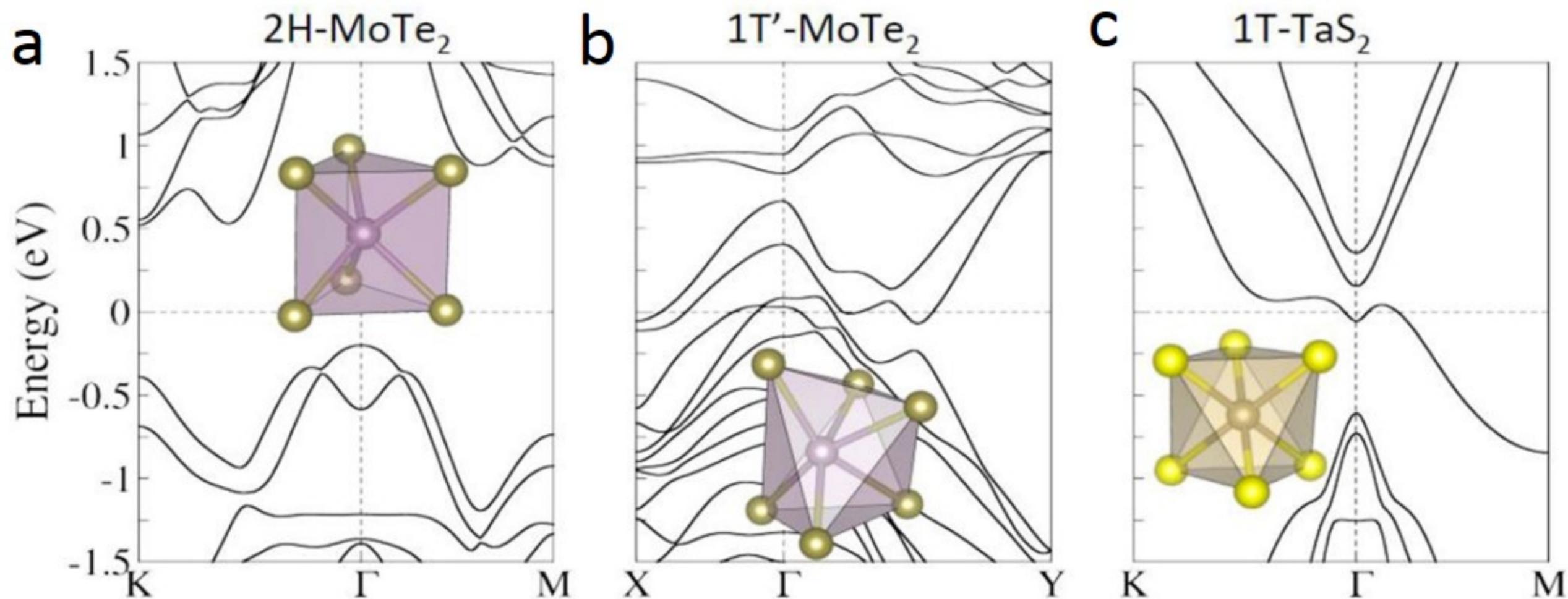
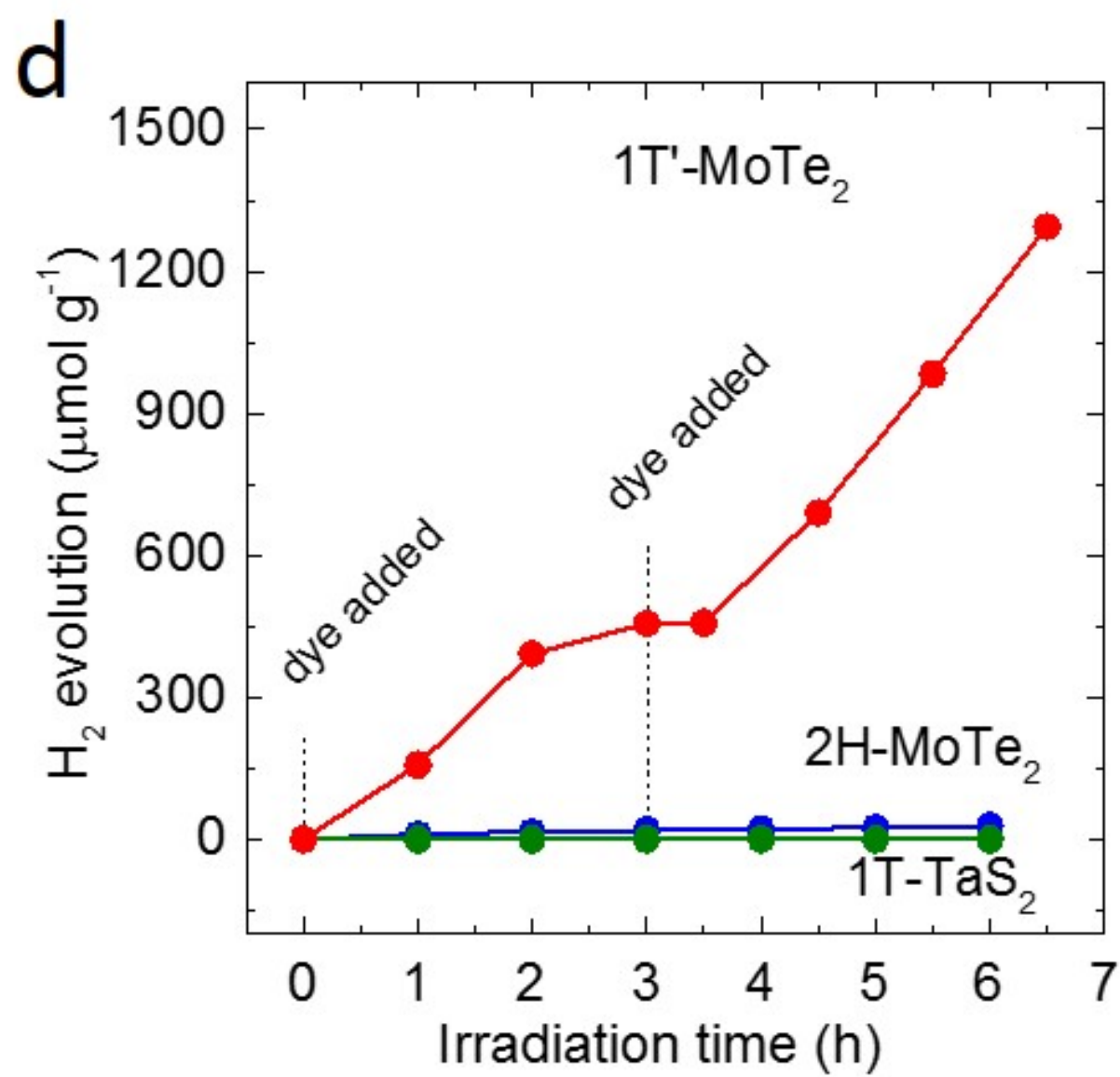
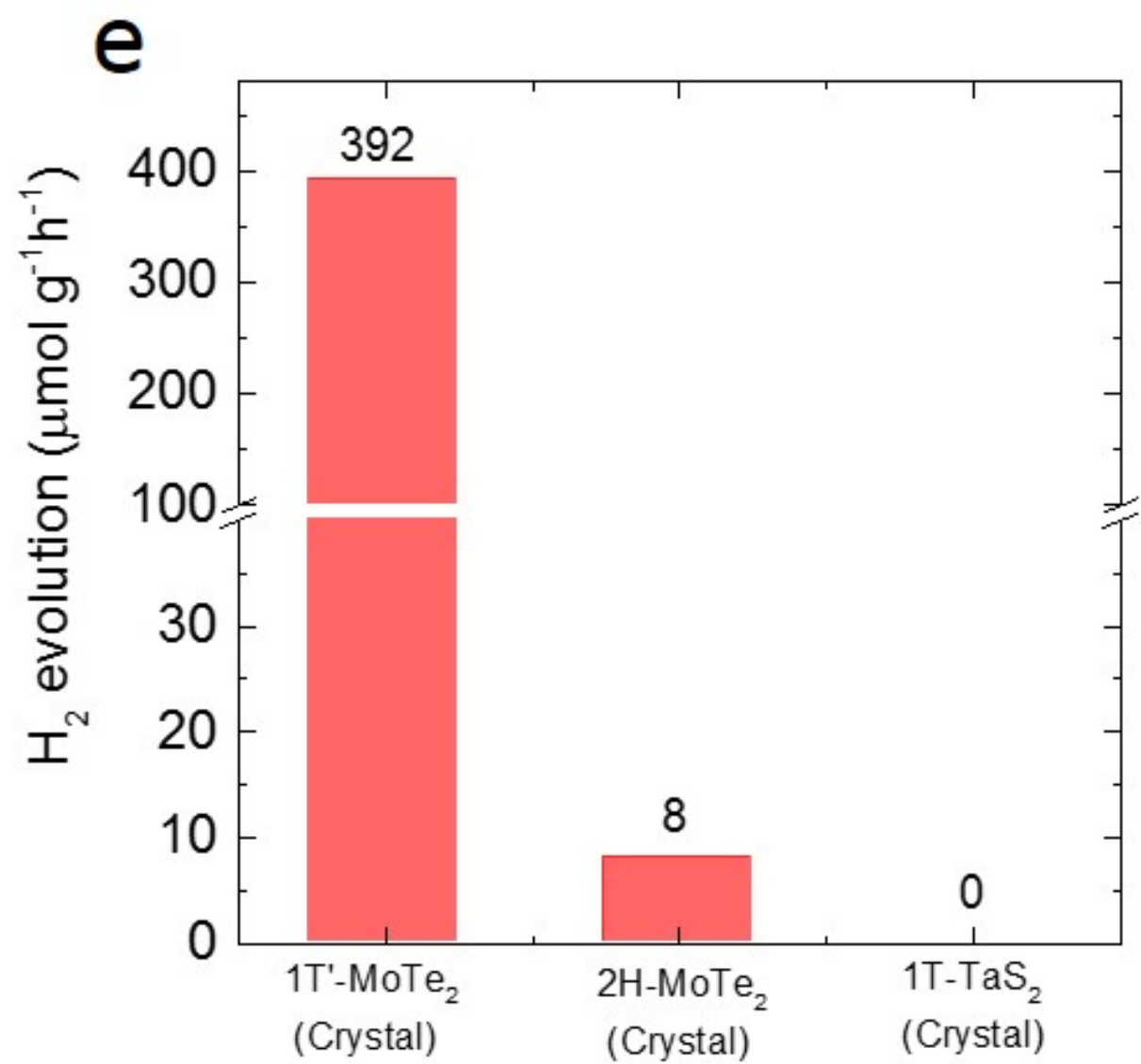

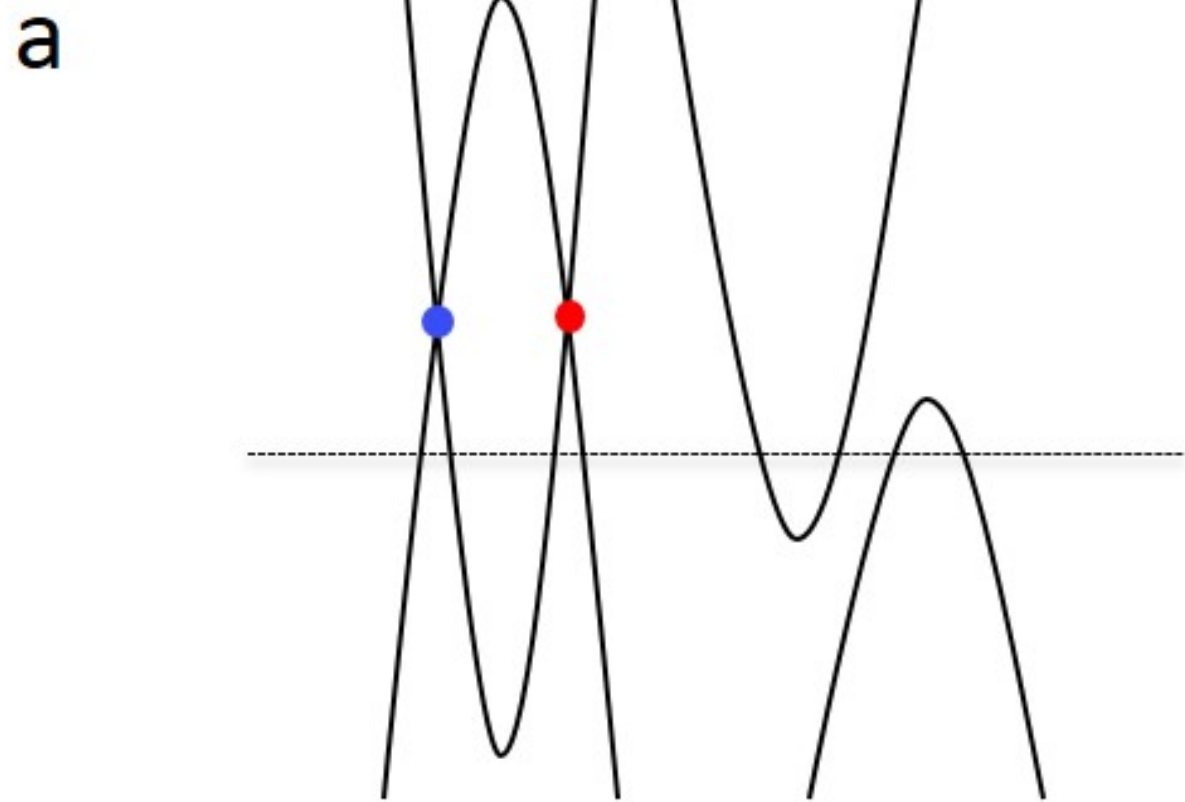
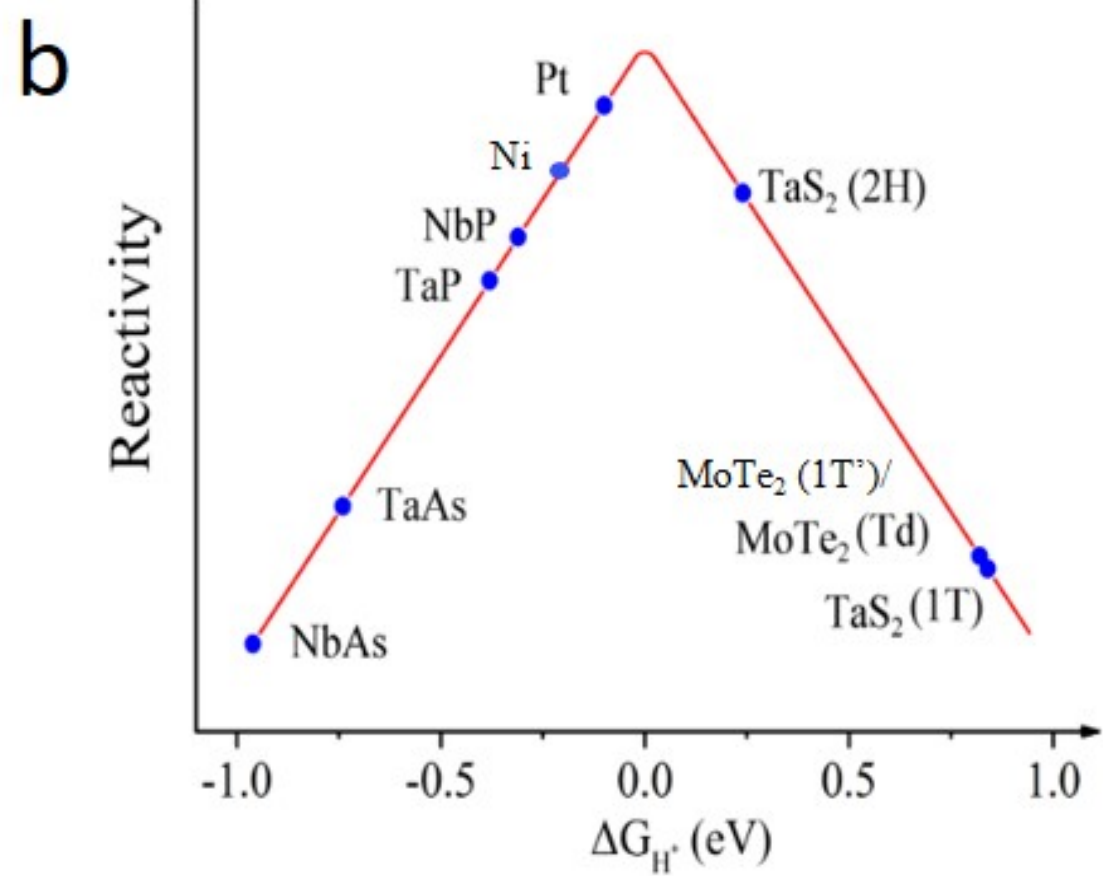
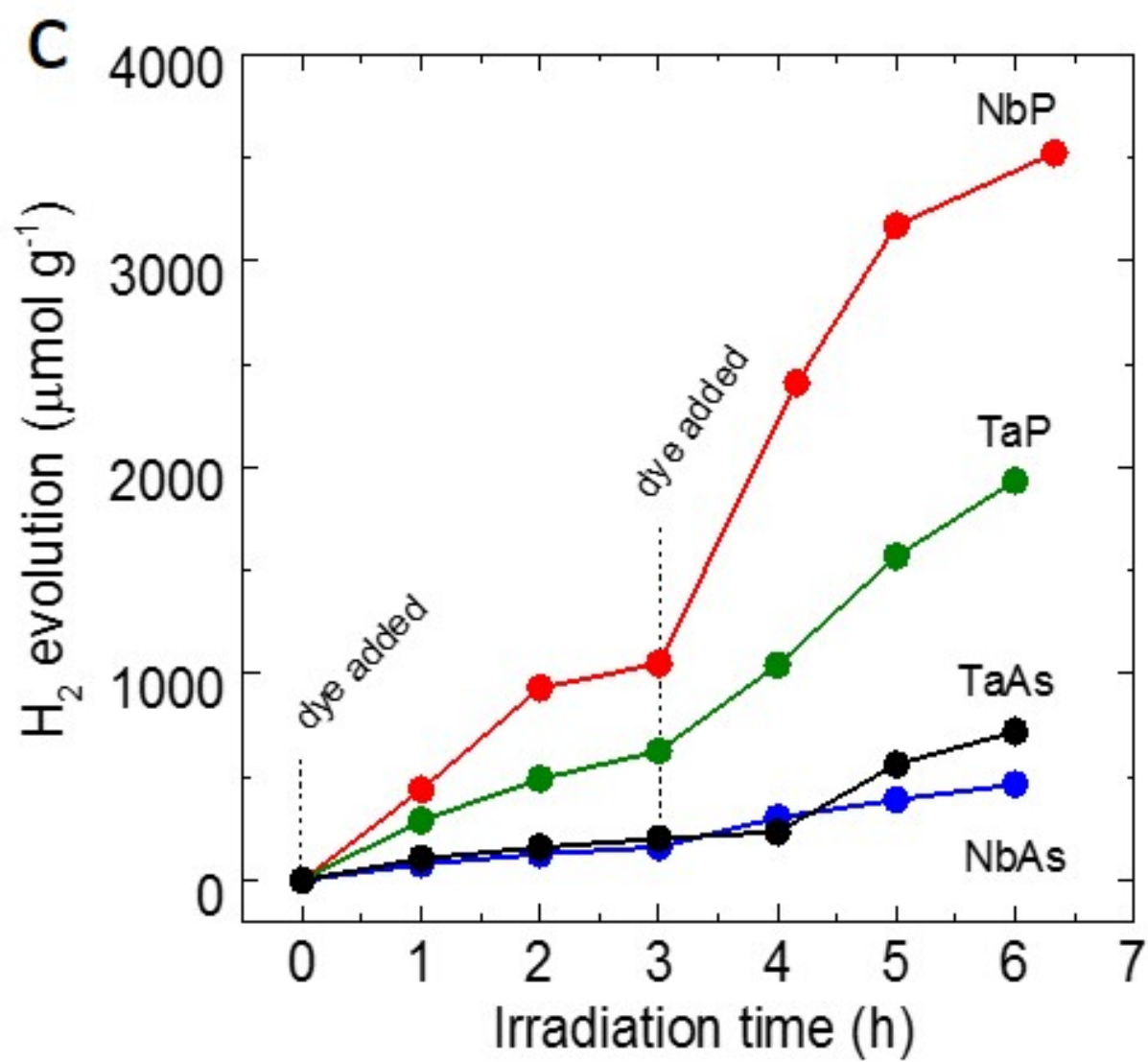
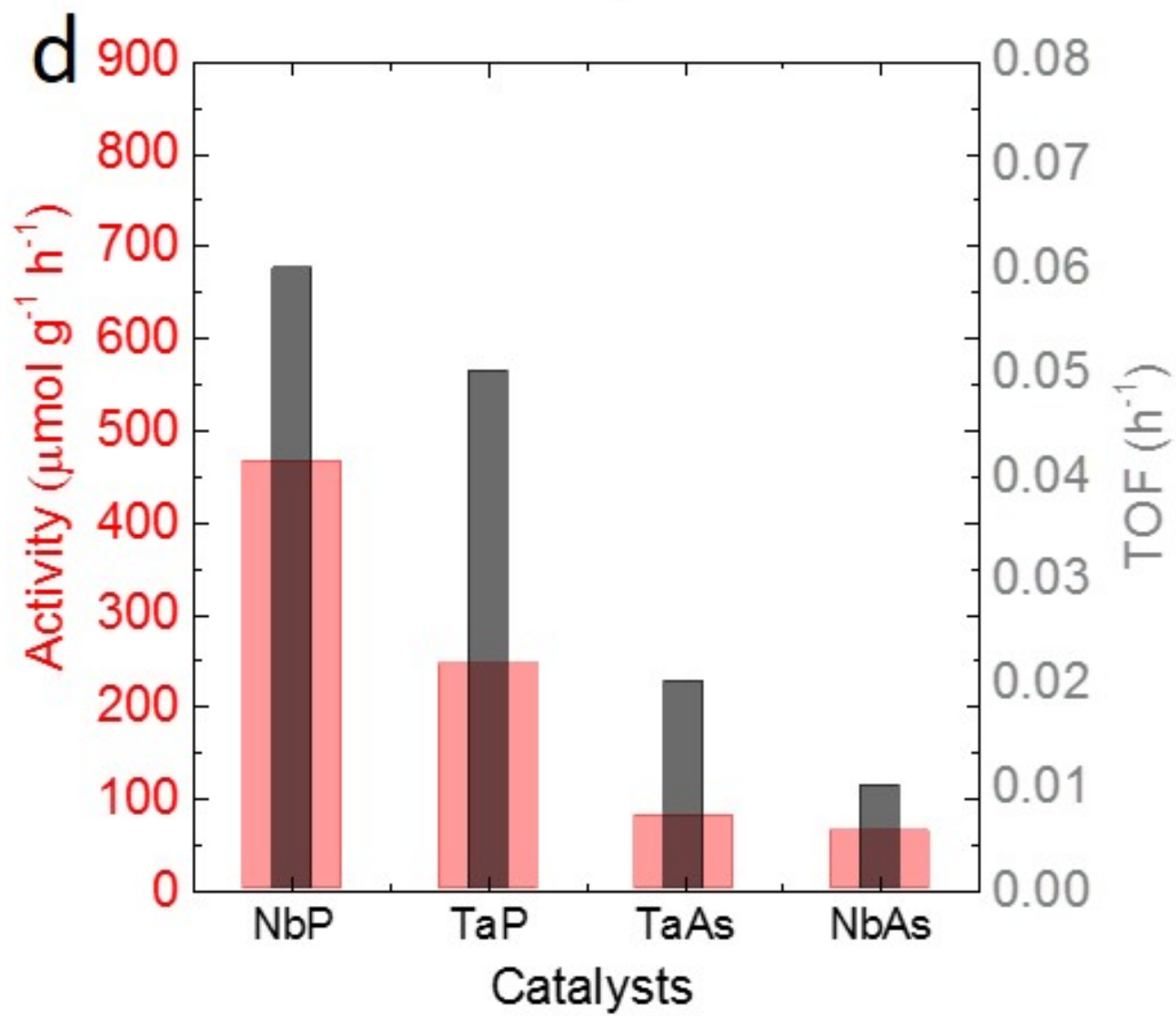

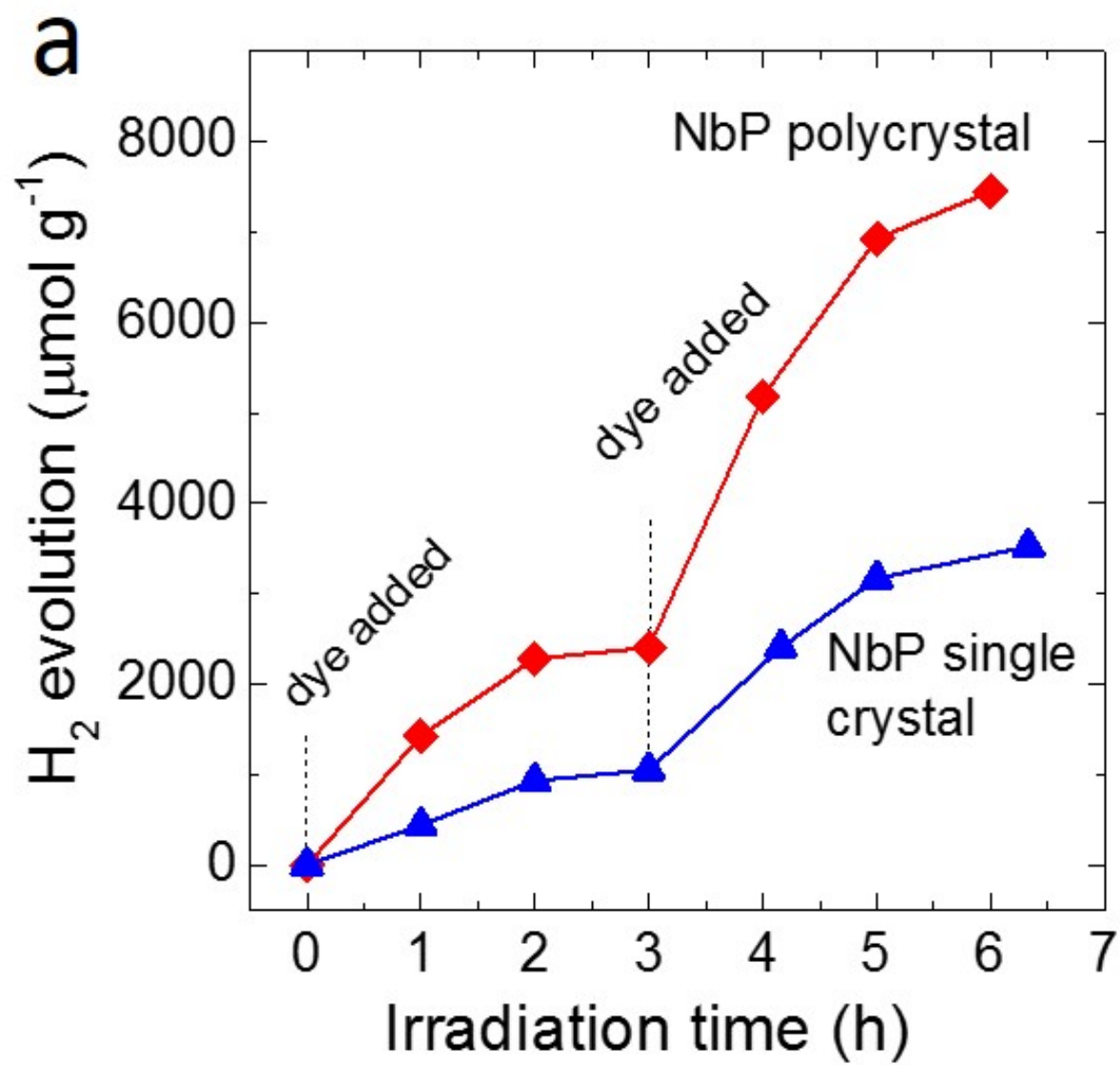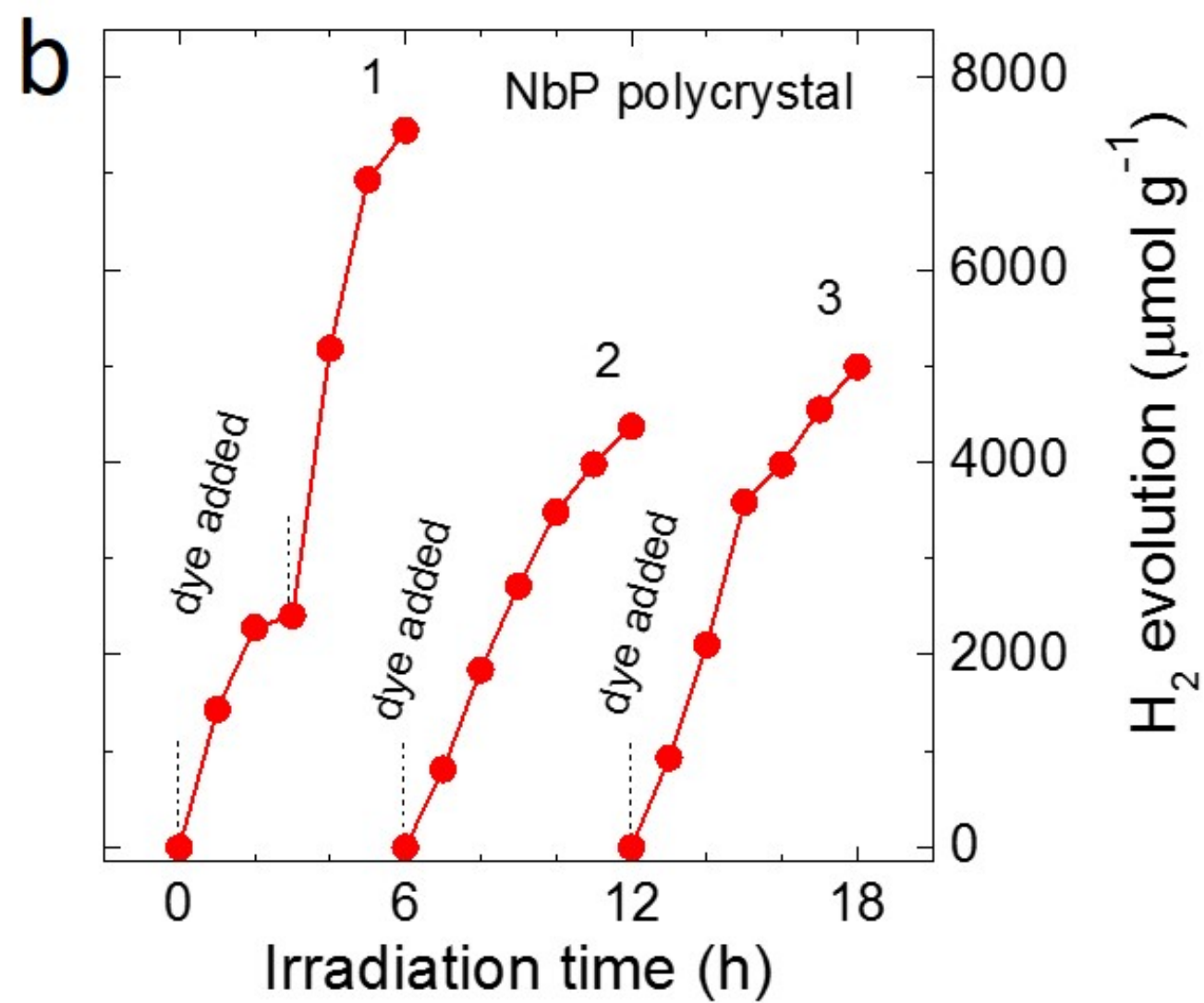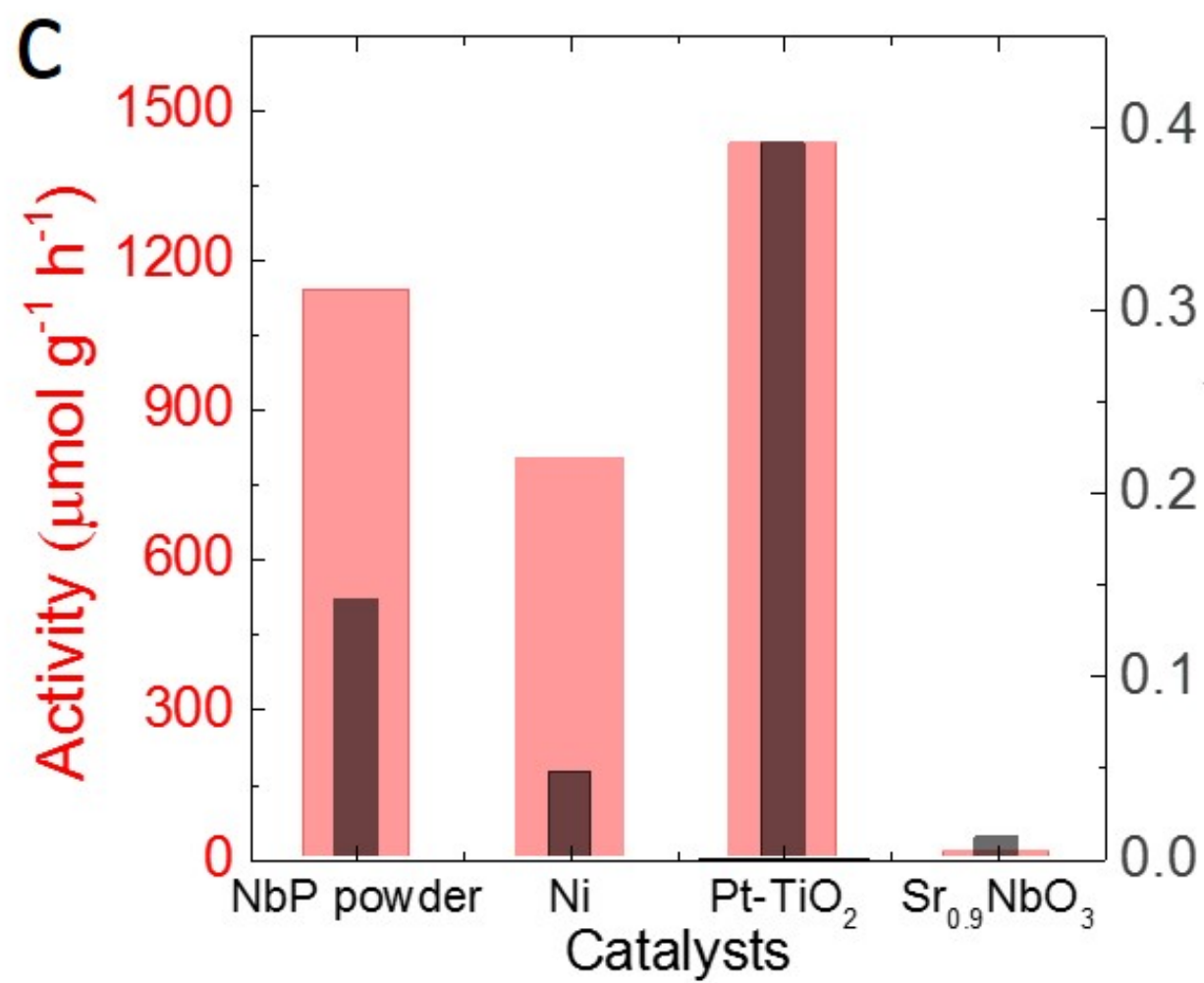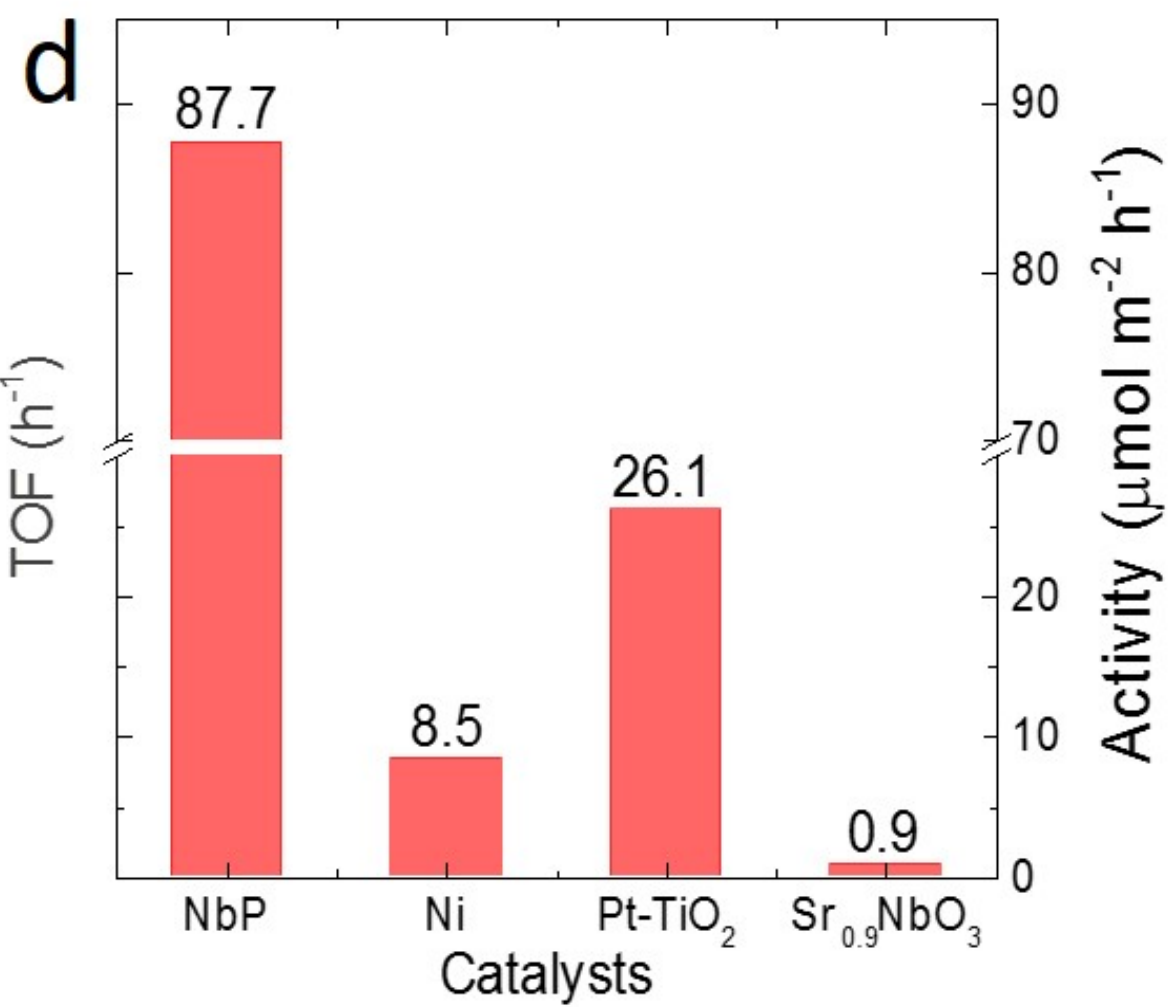